\documentclass[preprint]{aastex}
\input epsf
\newcommand{\reff}{\mbox{$r_{\rm e}$}}

\newcommand{\msun}{\mbox{M$_{\odot}$}}
\newcommand{\lsun}{\mbox{L$_{\odot}$}}
\newcommand{\mlv}{\mbox{$M/L_V$}}

\newcommand{\vi}{\mbox{$V\!-\!I$}}

\begin{document}
\title{Structural Parameters and Dynamical Masses for Globular Clusters in M33
 \footnote{Based on data obtained at the W.M. Keck Observatory, which is 
 operated as a scientific partnership among the California Institute of 
 Technology, the University of California and the National Aeronautics and 
 Space Administration.}
}

\author{S{\o}ren S. Larsen and Jean P. Brodie
  \affil{UC Observatories / Lick Observatory, University of California,
         Santa Cruz, CA 95064, USA}
  \email{soeren@ucolick.org and brodie@ucolick.org}
\and
  Ata Sarajedini
  \affil{Department of Astronomy, University of Florida, 
         211 Bryant Space Science Center, P. O. Box 112055,
	 Gainesville, FL 32611-2055, USA}
  \email{ata@polaris.astro.ufl.edu}
\and
  John P. Huchra
  \affil{Harvard-Smithsonian Center for Astrophysics, USA}
  \email{huchra@fang.harvard.edu}
}

\begin{abstract}
  Using high-dispersion spectra from the HIRES echelle spectrograph on the
Keck I telescope, we measure velocity dispersions for 4 globular clusters
in M33. Combining the velocity dispersions with integrated photometry and
structural parameters derived from King--Michie model fits to WFPC2 images, we
obtain mass-to-light ratios for the clusters. The mean value is
$\mlv = 1.53\pm0.18$, very similar to the \mlv\ of Milky Way and M31
globular clusters. 
%This implies that the M33 clusters have similar ages
%to Milky Way globular clusters, unless their stellar IMF is strongly biased
%towards low-mass stars.  
The M33 clusters also fit very well onto the fundamental plane and
binding energy -- luminosity relations derived for Milky Way GCs. 
Dynamically and structurally, the four M33 clusters studied here appear 
virtually identical to Milky Way and M31 GCs. 
\end{abstract}


\section{Introduction}

  In the Local Group, by far the best studied globular cluster (GC) systems 
are those of the Milky Way and M31. Significant amounts of data have been
gathered for globular clusters in these galaxies, with extensive compilations
in \citet{har96} (Milky Way, 147 clusters) and \citet{barm00} (M31, 435 
clusters).  Both of these large spirals are known to contain two 
distinct GC subpopulations: a metal-poor halo population and a more metal-rich 
population, associated with the bulge or thick disk 
\citep{kin59,zin85,min95,jab98,per02}. The presence of two GC subpopulations
presumably indicates different formation processes, and may hold clues to
the formation and early evolution of the host galaxies. A particularly
controversial issue is whether or not age differences exist between globular
clusters of different metallicities.  In M31, \citet{bh00} found evidence for
age differences of 4--8 Gyrs between the metal-rich and metal-poor clusters
and \citet{sara97} concluded that a similar age spread exists among GCs in 
the Milky Way. Other recent studies have proven less conclusive, but age
differences of a few Gyrs cannot be ruled out \citep{svb96,buo98,ros99}. 

  The GC system of the third spiral galaxy in the Local Group, M33, is
more sparse and consequently more difficult to identify and study.  Early 
work on the star clusters in this galaxy was done by \citet{hil60} and 
\citet{km60}, who noted that some of them had unusually blue colors. 
Photometry for about 130 cluster candidates was given by \citet{cs82,cs88},
who tentatively identified 27 old GC candidates in M33. They
also noted that M33 appears to contain a number of ``blue populous''
clusters similar to those in the Large Magellanic Cloud \citep{hod61},
but without the pronounced age gap that exists between $\sim3$ Gyrs and 12--15
Gyrs in the LMC \citep{gir95}. The old GCs exhibit halo-like kinematics,
while the younger clusters follow the rotation of the disk \citep{schom91}.
Color-magnitude diagrams for 10 halo GCs in M33, obtained with WFPC2 on board
HST, were presented by \citet{sara98}, who noted that most of the clusters
showed no blue horizontal branches in spite of having low metallicities. They
suggested that this might be due to a second-parameter effect, indicating
that many of the ``old'' M33 clusters may be several Gyrs younger than their
Galactic counterparts.  Additional clusters have been identified on WFPC2 
images by \citet{chan01}, and again \citet{chan02} suggested a larger age 
spread among the halo GCs compared to the Milky Way.

  It is important to establish 
how much the old globular clusters in different galaxies have in common.
A larger age spread among halo GCs in M33, for example, might indicate that
this galaxy assembled over a longer period of time, while the
structure of individual clusters may provide information about the
gas clouds out of which they formed
\citep{ml92,ee97}. In the Milky Way, a number of fundamental correlations
are known to exist between parameters such as surface brightness, core 
radius and velocity dispersion of globular clusters \citep{korm85,djorg95}. 
\citet{mac00} found that all Milky Way GCs are consistent with a
constant $V$--band core mass-to-light ratio (within the error margins) of 
1.45 $\msun \, \lsun^{-1}$. He also found a tight relation between the 
binding energy 
of individual clusters and the total luminosity, with a weak decrease in 
binding energy as a function of Galactocentric distance. The dependence
on Galactocentric distance may hold information about pressure gradients
in the early Galactic halo.

  Detailed information about the structure of extragalactic GCs, even
within the Local Group, is more limited for obvious reasons. At the
distance of M31 and M33, a typical ground-based resolution of $\sim1\arcsec$
corresponds to a linear scale of about 4 pc, roughly similar to the
typical half-light radii of globular clusters. Thus, information about the 
detailed structure of individual clusters is not easily obtained.
However, with the HST the situation is greatly improved, with
one pixel on the Planetary Camera (PC) corresponding to about 0.2 pc.
This is of great importance not only for studies aiming at a morphological
characterization of the clusters, but also for dynamical studies in 
which velocity dispersions must be tied to knowledge about the cluster
structure in order to provide accurate mass-to-light ratios. So far,
the mass-to-light ratios and other properties of globular clusters in M31
appear to be identical to those observed in Milky Way GCs 
\citep{djorg97,dg97,barm02}. Similarity in the M/L ratios (modulo metallicity
effects) is, of course, to be expected if the clusters have similar ages, unless 
there are variations in the stellar mass function. No information about dynamical 
masses has yet been published for M33 GCs, but could potentially be used as a tool 
to independently check whether age differences exist.

  In this paper we study 4 globular clusters in M33, selected from
the sample of \citet{sara98}. The clusters were originally selected based
on their halo-like kinematics and red colors ($\bv > 0.6$), and should be 
as close an analogy to the halo clusters in the Milky Way as possible. We 
use new high-dispersion spectroscopy from the HIRES spectrograph on the Keck I
telescope to measure velocity dispersions for the individual clusters, and
combine these with structural information from the same HST images used 
by \citet{sara98} to derive mass-to-light ratios and compare with data for
Milky Way globular clusters. Throughout the paper we will adopt a
distance modulus of 24.84 and the reddenings for each cluster determined by
\citet{sara00}.

%  Assume distance modulus of 24.84, from \citet{sara00}
%  Summary of Sarajedini work \citep{sara00}.
%   
%  Earlier work on Milky Way and M31 globulars \citep{djorg97,dg97}.
% 
%  \citet{mac00} and other papers \citep{korm85,djorg95} on GC fundamental plan.
% 
%  M33 clusters:
%
%  \citep{schom91} - Kinematics of 45 Star Clusters in M33. Red clusters
%   (B-V$>$0.6) have halo-like kinematics, with no system rotation.
%
%  \citep{chan99a} - 60 star clusters in 20 WFPC2 fields.
%
%  \citep{chan99b} - Physical properties -- fitted \citet{king62} models.
%    Cluster R12 is in common with our sample:
%    They find $r_c = 0\farcs234$ (0.96 pc, accounting for WFPC2 PSF)
%     or 1.08 pc (direct fit). We get $0\farcs225$ = 1.01 pc (assuming 
%     slightly larger distance) from \citep{king66} fits.
%
%  \citep{chan01} - detect 102 star clusters, of which 82 previously
%    unknown. Claim to discover 15 new GC candidates based on V-I colors.

\section{Data}

  Observations of the four clusters (R12, H38, M9 and U49) were carried out on 
Oct 24 and 25, 1998 with the HIRES spectrograph \citep{vogt94} on the Keck I 
telescope. A slit width of 0\farcs725 was used, providing a spectral 
resolution of $\lambda/\Delta\lambda \approx 54000$. The wavelength range
was 3730--6170 \AA, distributed over 38 echelle orders, although a useful 
S/N was achieved only for $\lambda \ga 4000$ \AA . For each 
cluster, 7--8 individual exposures of 1800 s each were obtained. In addition, 
the star HD 1918 (G9 III) was observed as a template star for the velocity 
dispersion measurements.

  Extraction of the spectra from the CCD images was done with
the highly automated MAKEE package, written by T.\ Barlow and specifically
tailored for reduction of HIRES data. MAKEE
automatically performs bias and flatfield corrections, cleans the images
of cosmic-ray (CR) hits and then locates and extracts each echelle 
order from the CCD images. Wavelength calibration was done using spectra of 
ThAr calibration lamps mounted within HIRES, with zero-point corrections 
based on sky lines.  The individual 1d spectra of each cluster were co-added 
using a sigma-clipping algorithm to eliminate any residual CR events,
producing a S/N for the co-added spectra of 15--25 per resolution element.

\subsection{Velocity dispersions}
\label{sec:vdisp}

  A variety of techniques have been developed to measure velocity dispersions
from integrated spectra in cases like ours where the expected dispersions 
(5--10 km/s) are comparable to the instrumental 
resolution. The most direct method is to simply convolve the
template star spectrum with Gaussian profiles corresponding to different
velocity dispersions, and then find the velocity dispersion that minimizes
the residuals when the (scaled and smoothed) template star spectrum is
subtracted from the cluster spectrum. One appealing aspect of this method
is that the quality of the fit and the match between template and
object spectra can be readily inspected, but the 
velocity dispersions are quite sensitive to an accurate determination of
the velocity difference between the two objects as well as a good match
between the cluster and template spectra. Alternatively,
the fitting can be done in the Fourier domain \citep{il76}, utilizing the fact
that the slope of the power spectrum is strongly sensitive to the spectral 
resolution.  A third method is to cross-correlate the cluster spectrum with 
the template spectrum and use the width of the cross-correlation peak as an
indicator of the velocity dispersion \citep{td79,hf96}. Yet another technique 
was used by \citet{dg97}, who convolved the cluster spectrum with a suitably
designed mask to obtain a ``mean'' spectral line whose width is related to
the velocity dispersion.

  The method of directly fitting a smoothed template spectrum to the
cluster spectrum is illustrated in Figure~\ref{fig:vd}, where we compare 
two echelle orders of the spectrum of cluster H38 with the template star. 
Each panel shows the cluster spectrum, the template star spectrum, the 
best-fitting smoothed template star spectrum and the residuals. Even though 
the smoothing results in a much better fit than the raw template spectrum, 
indicating that the broadening of lines in the cluster spectrum due to 
velocity dispersion is well resolved, significant residuals are still visible 
for many of the stronger lines.  There are no systematic trends in the
residuals (some are positive, others negative), but the template star is 
evidently not an ideal match to the cluster spectrum.

  We also measured velocity dispersions with the cross-correlation method,
using the FXCOR task in the RV package in IRAF\footnote{
  IRAF is distributed by the National Optical Astronomical Observatories, 
  which are operated by the Association of Universities for Research in 
  Astronomy, Inc.~under contract with the National Science Foundation
}. 
This method is less dependent on an exact match between the cluster and
template spectra, because a poorer match will tend to change the amplitude
rather than the width of the cross-correlation peak.  In Fig.~\ref{fig:ccf} 
we show the cross-correlation functions (CCFs) for cluster H38 (order 28) 
vs.\ the 
template star (solid line), as well as the CCFs for the template star spectrum 
broadened with three different Gaussians (0, 6 km/s, 12 km/s) vs.\ the 
template star spectrum itself (dashed lines).  The best fit is obtained for 
a velocity dispersion around 6 km/s. As a useful ``byproduct'', the 
cross-correlation technique automatically provides radial velocities for the 
clusters, which are needed as input for the other methods.

  Line-of-sight velocity dispersions ($\sigma_x$) are listed in 
Table~\ref{tab:rv} for both the direct fitting and cross-correlation 
techniques.  The $\sigma_x$ values are averaged over all fitted echelle 
orders, with the number of echelle orders used for the fits listed in column 
(2). We give both the rms scatter of the $\sigma_x$ values measured on the 
individual echelle orders, 
as well as the formal standard errors on the mean. However, many of the
uncertainties, related to choice of template star, fitting technique etc.\ 
are systematic rather than random, and simply estimating the uncertainties
from the scatter between the various echelle orders may not give a realistic 
estimate of the true uncertainties.  Table~\ref{tab:rv} also lists the radial 
velocities of the clusters and measurements from \citet{schom91} and 
\citet{chan02} for comparison.  The dominating uncertainty on our radial 
velocity measurements is actually the radial velocity of the template star 
itself, given as $+35.7\pm2$ km/s in SIMBAD.  For all clusters except H38, 
our radial velocities agree with those of \citet{chan02} within the errors.
The mean difference between our measurements and theirs is $19\pm13$ km/s.

  As seen in Table~\ref{tab:rv}, the cross-correlation technique gives 
systematically lower velocity dispersions than the direct fitting, with 
differences of 0.4 -- 0.9 km/s. We tested the direct fitting
method by artificially broadening a template star spectrum, adding
noise and applying the fitting algorithm to the resulting spectrum. 
The dispersions of the Gaussians used for the smoothing were reproduced with 
an accuracy of better than 0.1 km/s for a S/N similar to that of the
cluster spectra.  For even lower S/N the accuracy decreased, but no
systematic trends were evident.  However, we note that one pixel 
corresponds to a velocity difference of about 2.0 km/s, so
the difference between the two methods may well be due to centering and/or
binning effects.  Direct fitting is more sensitive to such effects than the 
cross-correlation technique and we may assume that the latter is more 
accurate.  In the following discussion we therefore adopt the velocity 
dispersions obtained by the cross-correlation method. 

\subsection{Structural parameters}
\label{sec:struct}

  To obtain structural parameters for the clusters we used 
images from the Wide Field Planetary Camera 2 on board the Hubble Space
Telescope, described in \citet{sara00}. The clusters are in all cases roughly
centered on the Planetary Camera (PC) chip. Each dataset contained exposures
in the F555W ($V$) and F814W ($I$) bands, allowing for two independent
sets of measurements for each cluster. 

  From visual inspection of the WFPC2 images (Figs 1, 2, 4 and 6 in
Sarajedini et al.\ 2000) it is clear that none of the clusters are
strongly elongated. We used the ELLIPSE task in the STSDAS package
in IRAF to fit elliptical isophotes to the cluster images, after filtering
the images with an $11\times11$ pixels median filter to create the smooth 
profiles required by ELLIPSE. Fig.~\ref{fig:efit} shows the position angle
(N through E) and ellipticity as a function of semi-major axis for each
cluster and confirms that all clusters have small ellipticities.  For radii 
less than about $0\farcs5$, the smoothing and random fluctuations due to 
individual stars make the fits meaningless. For H38 and M9, strong variations 
in the position angle are seen also at larger radii, most likely because the 
ellipticities are so small that the PA is very poorly constrained.  U49 is 
the only one of the four clusters where visual inspection of the images hints 
at some elongation, and the ELLIPSE fits yield an ellipticity of 
$\epsilon\approx$ 0.10--0.15 for this cluster, with a relatively stable 
PA around 30 degrees. For the remaining clusters the ellipticities are less 
than $\sim0.1$, consistent with the mean value of $0.07\pm0.01$ for Milky 
Way globulars \citep{ws87} and also in good agreement with the
mean $\epsilon=0.11\pm0.01$ for M31 GCs \citep{barm02}.  

  Additional structural parameters for the clusters were derived by fitting 
single-mass King--Michie models 
\citep[hereafter simply ``King'' models]{king66} 
to the WFPC2 images. Because the ellipticities
are generally small and the position angles ill-determined and possibly
varying with radius, we simply assumed circularly symmetric profiles
for the model fits. We then carried out a least-squares fit directly to the 
images, solving for the core radius $r_c$, concentration parameter $W$ and 
central surface brightness $\mu_0$. The minimization was done with the 
downhill simplex (``amoeba'') algorithm described in \emph{Numerical Recipes} 
\citep{press92}.  Although the clusters are quite well resolved, crowding 
near the center and strongly variable completeness functions as a function 
of radial distance made direct starcounts unfeasible.  For the same reasons, 
attempts to subtract individual stars from the images might also lead to 
systematic errors in the derived profiles.  Assuming that field stars are 
uniformly distributed throughout the frames, we decided that the approach 
that would be least likely to suffer from systematic effects due to partial 
resolution of the clusters was to simply fit the King models to the images 
directly.  The 
background level was determined as the average of all pixel values in an 
annulus centered on the cluster, with an inner radius of 200 pixels (9 arcsec) 
and 100 pixels wide. This is about the largest possible background annulus
that could be consistently used, considering variations in the exact position
of the clusters on the PC chip and the limited field size.
The fit itself was carried out for radii less than 200 
pixels.  We also attempted to solve for the centroid of the profile 
simultaneously, but these fits turned out to be unstable and we found that a 
more reliable approach was to determine the center of the profiles by 
carrying out the fits for various central positions and inspecting the 
residuals.  Even for fairly small centering errors ($\la1$ pixel), asymmetries 
in the residuals were clearly visible. 

  As noted by other authors \citep{chan01,barm02}, the effect of the HST
PSF is not entirely negligible when measuring structural parameters for
star clusters in M31 or M33. Our
modeling included a convolution of the 2-d King profiles with
the PC PSF, generated by the TINYTIM software\footnote{
  Available at the URL http://www.stsci.edu/software/tinytim/tinytim.html
}
and, like previous studies, we found small but measurable differences
in the fits when the PSF convolution was included. Core and effective
radii were typically $\sim0.1$ pc ($\sim0\farcs02$) smaller, while the 
central surface brightnesses $\mu_0$ increased by 0.1--0.2 mag with
PSF convolution.

  Fits to the F555W images are shown in Fig.~\ref{fig:kfitv}. The error
bars indicate the actual cluster profiles while the solid lines are the
King model fits. The light profiles were measured in concentric apertures 
with the PHOT task within DAOPHOT in IRAF, measuring the sky background in 
the same annuli that were used for the fits.  Note that, because of the 
2-d fitting procedure, the solid lines are \emph{not} direct fits to the actual 
datapoints in Fig.~\ref{fig:kfitv}, but are measured on the best-fitting 
convolution product of a 2-d King profile and the PC PSF, using the same 
procedure as for the cluster profiles.  The King models generally provide 
excellent fits to the globular cluster profiles.  In particular, we see no 
evidence for central cusps that might indicate core-collapsed clusters.
The fitted parameters are listed in Table~\ref{tab:kfit}, with separate fits 
for the F555W and F814W images.  Uncertainties were estimated by repeating 
the fitting procedure 18 times using different initial values for $r_c$, $W$ 
and $\mu_0$.  We have adopted the standard deviation on each fitted parameter, 
rather than the standard error on the mean, as our uncertainty estimates.  
Alternatively, the difference between the F555W and F814W fits may be used
as an estimate of the uncertainties, and provides similar uncertainty 
estimates to those in Table~\ref{tab:kfit}.

  Cluster R12 was also measured by \citet{chan99}. Although they fitted
the empirical version \citep{king62} of the King models rather than the 
theoretical King--Michie models used here, the two types of King models 
are structurally similar enough that a comparison of the derived core radii is 
illustrative. Fitting a \citet{king62} profile to the F555W image, 
Chandar et al.\ found a core radius of $0\farcs234$ for R12 which is only 
slightly larger than our value of $0\farcs225$ (corrected for the PSF). Without
correction for the PSF we get $r_c = 0\farcs253$, now slightly larger than
the Chandar et al.\ value. The small differences between our measurements and 
those of Chandar et al.\ can easily be ascribed to the different choices of 
King profiles and fitting procedure, small centering errors etc.

%  \citep{chan99b} - Physical properties -- fitted \citet{king62} models.
%    Cluster R12 is in common with our sample:
%    They find $r_c = 0\farcs234$ (0.96 pc, accounting for WFPC2 PSF)
%     or 1.08 pc (direct fit). We get $0\farcs225$ = 1.01 pc (assuming 
%     slightly larger distance) from \citep{king66} fits.

  In addition to the three basic parameters $r_c$, $W$ and $\mu_0$, 
Table~\ref{tab:kfit} also
includes the parameter $C$, which is the ratio of the tidal radius $r_t$ to 
the core radius $r_c$, as well as the effective radius $\reff$, defined as 
the radius containing half the cluster light in projection.  These two 
additional parameters both follow directly from the King model $r_c$ and 
$W$ parameters, but are listed for convenience. 

  One potential concern is that the tidal radius for one of the clusters,
U49, is larger than the inner radius of the sky annulus, leading to possible
systematic errors in the sky background determination. For the best-fitting 
model, a sky 
background level of about 0.2 ADU is expected at $r=200$ pixels, decreasing 
rapidly outwards. We tried varying the background level by $\pm0.5$ ADU and 
redid the fits for U49. The core and effective radii were affected by less 
than $0\farcs05$ (0.2 pc) while the dimensionless $C$ parameter changed by 
about 1, i.e.\ the changes were in both cases comparable to the 
uncertainties in Table~\ref{tab:kfit}.

  The core and effective radii for the M33 clusters are well within the 
range spanned by Milky Way globular clusters. The $r_c$ distribution for 
Milky Way GCs peaks at about 1 pc, but has a tail extending up to above 
10 pc. Excluding core-collapsed clusters, the median value is 1.3 pc. The 
median half-light radius \reff\ for Milky Way GCs is 3.3 pc \citep[using 
values from][]{har96}, but the distribution is again strongly asymmetric. 
It may be worth noting that the reddest M33 cluster 
(R12, see sect~\ref{sec:phot}) also has the smallest
effective radius, an effect that has been observed in many other galaxies
including the Milky Way and M31 \citep[e.g.][]{barm02,lar01}.
Another point of interest is that previous studies of Milky Way and M31
globular clusters have found that low-concentration clusters tend to
be more elliptical \citep{barm02}, which is consistent with U49 being the 
most elongated as well as the largest of the clusters in our sample.  
However, the 4 clusters studied here clearly constitute a too small sample
to draw conclusions as to whether or not the same effects are generally 
present in M33.

\subsection{Integrated photometry}
\label{sec:phot}

  Fig.~\ref{fig:cgrow} shows the integrated $V$ magnitudes of the clusters
as a function of radius. The curves-of-growth are drawn with dashed lines
beyond the tidal radii obtained from the King profile fits. The background 
level was determined in the same way as for the King profile fits and the 
number counts in F555W and F814W in each aperture were converted to $V$ 
magnitudes and \vi\ colors using the
transformations in \citet{hol95}. Because the Holtzman et al.\ zero-points
include an implicit $-0.1$ mag aperture correction from their $r=0\farcs5$ 
reference aperture to infinity, which does not apply in our case, we have 
added 0.1 mag to the magnitudes. Apart from small color terms, this
gives exactly the same results as using the photometric zero-points in the 
image headers directly.
  
  The curve-of-growth for M9 has a ``bump'' at $r=200$ pixels, due to a bright 
star that is located just within the outer aperture radius, but the curves
are otherwise smooth and monotonically increasing over most of the radial
range. For a couple of clusters (R12 and H38), the curves decrease slightly
at large radii, reflecting the inherent uncertainties in the background
determination.  A small decrease of 0.5--1.0 ADU in the background levels
would make the curves-of-growth for R12 and H38 increase monotonically out 
to their tidal radii.

  Both \citet[][hereafter CS88]{cs88} and \citet[][CBF01]{chan01} have
published integrated photometry for the 4 clusters, using ground-based
CCD imaging and WFPC2 data, respectively.  In Table~\ref{tab:phot} we compare 
our photometry with CS88 and 
CBF01.  CS88 used an aperture radius of $3\farcs7$, corresponding to 82 pixels
on the PC camera, while CBF01 used an aperture of $2\farcs2$ (48 pixels)
for H38, M9 and R12 and $2\farcs7$ (59 pixels) for U49.  
% From 
% Fig.~\ref{fig:cgrow} one can see that these apertures only include a fraction 
% of the total flux, and it is somewhat surprising that CBF01
% get almost exactly the same $V$ magnitudes as CS88 in spite of using a 
% significantly
% smaller aperture.  
We redid the photometry in the same apertures used by
CBF01 and CS88 as well as in a larger aperture. For R12 and H38 this larger
aperture coincided with the tidal radii of the clusters, for M9 we used
an aperture of $r=190$ pixels (to avoid the bright star) and for U49 we 
used $r=200$ pixels.  For the comparison with CBF01 we used the same
background annulus as they did ($3\farcs5<r<5\arcsec$), but for the other
measurements the same annulus was used as for the King model fits and the
curves-of-growth.  

  While our \vi\ color measurements agree fairly well with those of CBF01, our 
$V$ magnitudes are systematically fainter than theirs by about 0.18 mag.  
Part of this discrepancy is due to the fact that no correction was applied
to the Holtzman et al.\ zeropoints in CBF01 (Chandar, priv.\ comm.). 
By experimenting with the PHOT options, we found that the remaining 0.08 mag 
can be accounted for by differences in the background level estimates.
Because the resolution into individual stars makes the sky histogram highly 
asymmetric, the background measurements in these fields are very sensitive 
to the choice of algorithm. Specifically, if we use the MODE option (as did 
CBF01) instead of simply calculating the 
mean, the discrepancy is reduced to a few times 0.01 mag. However, because 
the contribution from individual, resolved stars is part of the ``sky'' 
background within the photometric apertures and therefore needs to be 
subtracted, a simple mean appears to be the most appropriate background 
estimator for our purpose.

  Our photometry in the $3\farcs7$ aperture agrees closely with that of CS88, 
being only 0.02 mag fainter on the average.  For some clusters the 
curves-of-growth continue to rise well beyond $3\farcs7$, and the measurements
in Table~\ref{tab:phot} suggest that this aperture may underestimate the
luminosity of the most extended cluster, U49, by $\sim0.2$ mag. However, 
measurements in the largest apertures are quite sensitive to 
uncertainties in the background level, and for R12 and H38 the integrated 
magnitudes in the largest aperture are actually slightly fainter than in the 
$3\farcs7$ aperture, in accordance with the behavior of the curves-of-growth 
pointed out above. The small adjustments in the
background level that would make the curves-of-growth monotonic would affect 
the photometry at $r=50$ pixels ($2\farcs3$) by only $\sim0.02$ mag, but at
the tidal radii the changes amount to 0.1--0.2 mag.  This may provide a 
more realistic estimate of the true uncertainty on the integrated photometry
than pure photon statistics, which give formal errors of less than 0.002 mag.
In practice, the uncertainty on the background level is dominated by
stochastic variations in the numbers of individual, bright stars within the
aperture annulus, as well as true variations due to dust obscuration
etc. A larger field of view would help reduce the stochastic variations,
but would increase the sensitivity to large-scale variations in the
background.  For the following discussion we will use our photometry 
in columns (10) and (11) of Table~\ref{tab:phot} as the best approximation
to the total integrated magnitudes of the clusters.

\section{Masses and mass-to-light ratios}

  Once the velocity dispersion and structural parameters of a cluster
are known, the total mass can be estimated. The simplest way is to 
use the virial theorem which implies
\begin{equation}
  M_{\rm vir} \, = \, a \frac{\sigma_{3D}^2 r_h}{G}
  \label{eq:mvir}
\end{equation}
where the constant $a$ has a value of 2.5 \citep[][p.\ 11]{spi87}. 
In this expression, $\sigma_{3D}$ is the three-dimensional velocity dispersion,
$\sigma_{3D}^2 = 3 \, \sigma_x^2$, and $r_h$ is the 3-dimensional 
half-\emph{mass} radius which is generally larger than the projected
half-\emph{light} radius \reff\ by a factor of 4/3. In the present case, however, the 
King model fits allowed $r_h$ to be directly evaluated for each individual 
cluster, without using \reff\ as an intermediate step.  An alternative 
method is to use ``Kings formula'' \citep{rt86,qdp95},
\begin{equation}
  M_{\rm king} \, = \, \frac{9}{2 \pi G} 
                       \frac{\mu r_c \sigma_0^2}{\alpha \, p}
  \label{eq:mking}
\end{equation}
  Here $\sigma_0$ is the \emph{central} (projected) velocity
dispersion and $r_c$ is the core radius of the cluster. The remaining
parameters, $\mu$ (a dimensionless mass, not to be confused with the
central surface brightness $\mu_0$), $p$ (ratio of central surface
density to central 3-d density) and $\alpha$ (ratio of half-width at
half maximum to core radius, usually close to 1) are all functions of
the concentration parameter and were again evaluated using the King model fits.

  The velocity dispersion $\sigma_{3D}$ used in (\ref{eq:mvir}) is the mean 
value averaged over the entire cluster, while the $\sigma_0$ used in 
(\ref{eq:mking}) is the central value. Neither of these is the actual observed 
value $\sigma_x$, since 
a slit of finite size was used for the observations.  Using the King model 
fits, which provide the velocity dispersion in the cluster as a function of 
radius, we estimated the aperture corrections from our slit to the central 
and global values, assuming a slit width of 0.72\arcsec\ (3.2 pc at 
the distance of M33) and infinite length.  The aperture corrections 
are relatively modest, generally amounting to less than 10\% in either 
direction. 
% and that the ratio $\sigma_\infty / \sigma_0$ is a function of 
% the concentration parameter only.

  It should be noted that both methods implicitly assume a constant M/L ratio 
as a function of radius and isotropic velocity distributions. In practice, 
some mass segregation is expected because of equipartition of energy, and 
has indeed been detected observationally in some cases 
\citep[e.g.][]{king95,fer97,sosin97}.  The King models used here are 
single-mass models and certainly an oversimplification of the true dynamical 
situation in the clusters, but in practice the difference between dynamical 
masses obtained by fitting single-mass King models and more sophisticated 
methods tends to be relatively small \citep[e.g.][]{djorg95,mey01}. 
Anisotropy is expected to occur mostly in the outer parts of the clusters
\citep{gg79,lup87,tak00} and should not have a strong impact on M/L ratios 
derived from the integrated light.

  The computed central and global ($\sigma_{\infty}$) velocity dispersions 
are listed in 
Table~\ref{tab:mass}, col.\ (2) and (3). Columns (4) and (5) give the 
corresponding King and virial masses, using the structural parameters
measured on the F555W frames. Considering that neither the rms nor 
the standard errors on the mean are likely to provide realistic estimates 
of the true errors on the velocity dispersions, we have adopted 0.5 km/s 
for the errors on $\sigma_x$, 
corresponding roughly to the typical difference between the measurements 
based on direct fit and cross-correlation. We have omitted errors on 
$\sigma_\infty$ and $\sigma_0$ in Table~\ref{tab:mass} since these scale, 
at least to first order, with the errors on $\sigma_x$. Column 
(6) lists the central density in $\msun \, {\rm pc}^{-3}$ which, per 
definition, is $\rho_0 \, = \, M_{\rm King} / (\mu \, r_c^3)$. The
$V$-band (King) mass-to-light ratios, \mlv , are given in col.\ (7) and
column (8) lists the binding energy of each cluster, calculated as in
\citet{mac00}.  For the mass-to-light ratios we have used the $V$ magnitudes 
in Table~\ref{tab:phot} and the distance modulus of 24.84 and the 
reddenings of individual clusters given in \citet{sara00}.  Other estimates 
of the distance modulus of M33 are $24.52\pm0.14$(random)$\pm0.13$(systematic)
\citep[][based on Cepheids]{lee02}, 
$24.81\pm0.04$(random)$\pm0.05$(systematic)
\citep[][red giants]{kim02}, and
$24.64\pm0.09$ \citep[][Cepheids]{fwm91}. The derived M/L ratios are 
inversely proportional to the assumed distance, and could be 
underestimated by up to 14\% if the smallest of the above distances is correct.

  As seen from Table~\ref{tab:mass}, the virial and King masses are generally
very similar. This agreement is not entirely fortuitous, since the two 
estimates are not independent. In fact, the ratio $M_{\rm King} / M_{\rm vir}$ is
\begin{equation}
  \frac{M_{\rm King}}{M_{\rm vir}} \, \propto \,
  \frac{\mu}{\alpha \, p} \, \frac{r_c}{r_h} \, \frac{\sigma_0^2}{\sigma_{3D}^2}
  \label{eq:ratio}
\end{equation}
where the constants have been excluded.  Since all the factors in Eq.~\ref{eq:ratio} 
are functions of the concentration parameter only, this is also true for 
$M_{\rm King} / M_{\rm vir}$.  This ratio is plotted in Fig.~\ref{fig:mcmp} 
and is evidently very close to 1 for the relevant range of concentration parameters.

  The average of the \mlv\ ratios in Table~\ref{tab:mass},
$\langle\mlv\rangle \, = 1.53\pm0.18$, is in excellent agreement with 
McLaughlin's $1.45\pm0.1$ value for the core \mlv\ of Milky Way GCs. 
Cluster R12 formally has a somewhat lower \mlv\ than the other three
clusters, but we note that this is also the most compact of the clusters
and it is possible that the effects of mass segregation and departures from 
the single-mass King model approximation are more pronounced here.
For 6 clusters in M31 with HST imaging, \citet{dg97} obtain a mean \mlv\
ratio of $\langle\mlv\rangle \, = 1.2\pm0.21$ or
$\langle\mlv\rangle \, = 2.0\pm0.44$, based on King model fits and use
of the virial theorem, respectively. These estimates nicely bracket our
mean value for the M33 clusters.  If we had used the velocity dispersions 
based on the direct fitting instead of those obtained by cross-correlation, 
the masses (and resulting \mlv\ ratios) would have increased by $\sim20$\%, 
still in good agreement with the Milky Way value. 

%  From HST WFPC2 color-magnitude diagrams, \citet{sara98} found red horizontal 
% branch morphologies for 8 out of 10 clusters, in spite of low metallicities. 
% Assuming that cluster age is the second parameter determining horizontal 
% branch morphology, they thus suggested that these clusters might have 
% intermediate ages.  Of the clusters studied in the present paper, only M9 has 
% a ``normal'' blue$+$red horizontal branch morphology for its metallicity, 
% while the rest have exclusively red HBs, according to \citet{sara98}.  
% \citet{chan02} attempted to constrain cluster ages using Balmer line 
% strengths and population synthesis models and quote ages of 12--17 Gyrs 
% and $\sim5$ Gyrs for U49 and H38, respectively.
%

%  Age (Gyr)  Log(Age)        Vmag(M0)      M/L (rel)
%                          MSCA     SALP   MSCA   SALP
%     15       10.18      6.8543   7.1739  1.00   1.00
%     12       10.08      6.5783   6.9678  1.29   1.21
%     10       10.00      6.3684   6.8113  1.56   1.40
%      5        9.70      5.7310   6.2746  2.81   2.29
%      2        9.30      4.7377   5.3953  7.02   5.15

\section{Fundamental plane relations}

  We finally explore how the M33 clusters fit into the ``fundamental plane''
relations obeyed by Milky Way globular clusters. Figure~\ref{fig:fp}
shows our M33 data compared with data for Milky Way GCs, taken from
\citet{pm93} and \citet{har96}. Core-collapsed Milky Way clusters have been
omitted from the plots. Panels (a) and (b) show the bivariate core parameter
relations \citep{djorg95}, with Milky Way GCs plotted as $+$ symbols
and M33 clusters as filled circles. As shown by Djorgovski, both of these
relations are consistent with GCs being virialized systems with a constant
$M/L$ ratio. The M33 clusters fit remarkably well onto the the fundamental 
plane relations obeyed by the Milky Way clusters, consistent with their similar 
mass-to-light ratios. We note that \citet{barm02} found slight deviations 
between the fundamental plane relations for their sample of M31 globular 
clusters and the Milky Way data, but it is unclear to what extent those 
differences are due to systematic measurement errors or real physical 
differences between the GCs in the two galaxies. 

  Panel (c) shows the cluster binding energies as a function of absolute 
$V$ magnitude, $M_V$. \citet{mac00} found a tight relation between these
two quantities for Milky Way GCs, with an additional weaker dependence on 
Galactocentric distance $r_{\rm gc}$. He gave the relation as
\[
\log (E_b / {\rm ergs}) \, = \, 
  [(39.86\pm0.40) - 0.4 \log(r_{\rm gc}/8 {\rm kpc})] \,
 + \, (2.05\pm0.08)\log(L/L_{\odot}). 
\] 
  For the four M33 globular clusters we find a mean value of
$\langle\log (E_b \, L^{-2.05})\rangle \, = \, 39.81$, with a scatter of 
only 0.13 dex.  Ignoring the term in $r_{\rm gc}$, we thus find once again that 
the M33 globular clusters fall nearly exactly on the same relation as the 
Milky Way globular clusters.

  The similarity to Milky Way GCs in general, and in the mass-to-light 
ratios in particular for M33 clusters is interesting in the context of the 
results by \citet{sara98} and \citet{chan02}, since a younger age should 
generally imply lower M/L ratios unless there is an excess of low-mass stars.  
For a Miller-Scalo IMF, for example, population synthesis models by Bruzual 
\& Charlot (priv.\ comm.) predict an increase in \mlv\ of about a factor of 
2 from 5 Gyrs to 12 Gyrs, and a factor $\sim5$ from 2 Gyrs to 12 Gyrs.  
Indeed, \citet{fis92} find a much lower M/L ratio ($\mlv \la 0.2$) for the 
$\sim2$ Gyr old cluster NGC~1978 in the LMC than for any of the clusters 
studied here.  However, a younger age for the M33 clusters is still possible 
if a steeper IMF slope conspires with a young age to produce the same M/L.  
Of our four GCs, \citet{chan02} list spectroscopic age estimates for U49 
(12--17 Gyrs) and H38 (5 Gyrs), but we note that absolute ages based on 
integrated spectra are still uncertain \citep[e.g.][]{schia02}.  On the other 
hand, if H38 and U49 are 
both old clusters ($\ga10$ Gyr), then alternative explanations will be 
required for their excessively red horizontal branch morphologies, the 
presence of stars above their RGB tips, and their anomalously bright red clump 
magnitudes \citep{sara98,sara00}.  Clearly, more work needs to be 
done before the question of a real, large age spread among M33 halo GCs can 
be answered with confidence.

\section{Summary}

  We have measured velocity dispersions for 4 old ``halo'' globular clusters
in M33 and combined these with structural parameters from WFPC2 images to
obtain mass-to-light ratios for the clusters. Our analysis shows that the
M33 GCs have \mlv\ ratios essentially identical to those of GCs in the
Milky Way and M31, as well as similar structural parameters (core- and
effective radii). The M33 clusters also fall on the same 
``fundamental plane'' and binding energy vs.\ luminosity relations as 
Milky Way and M31 clusters \citep{djorg95,dg97,mac00}. Unless the stellar
IMFs in these clusters have an excess of low-mass stars relative to
GCs in the Milky Way, the similarity in M/L ratios and the fundamental
plane relations suggest similar high ages.  It would be very interesting to 
measure velocity dispersions for M33 clusters spanning a wide range of ages 
and look for the expected variations in M/L ratio.

\acknowledgments

  This work was supported by National Science Foundation grant number 
AST9900732 and by NSF Career Grant No.\ AST0094048. JPH was supported by
the Smithsonian Institution.  This research has made use of the SIMBAD 
database, operated at CDS, Strasbourg, France. We thank Rupali Chandar 
for helping to sort out the differences between our photometry and hers,
and the Keck staff for their assistance with the observations.

\newpage
% \onecolumn

\begin{flushleft}
\epsfxsize=12cm
\epsfbox{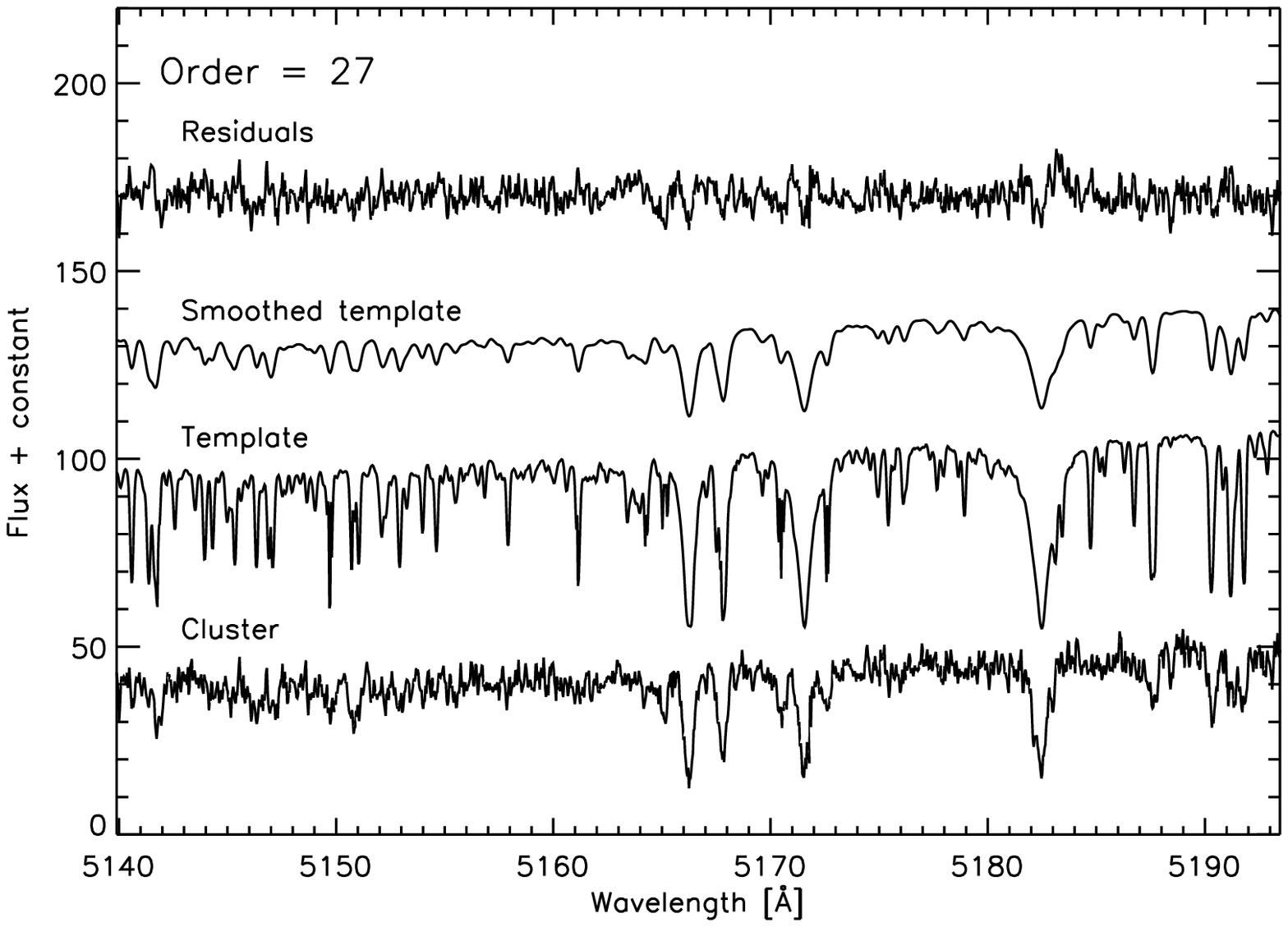}

\epsfxsize=12cm
\epsfbox{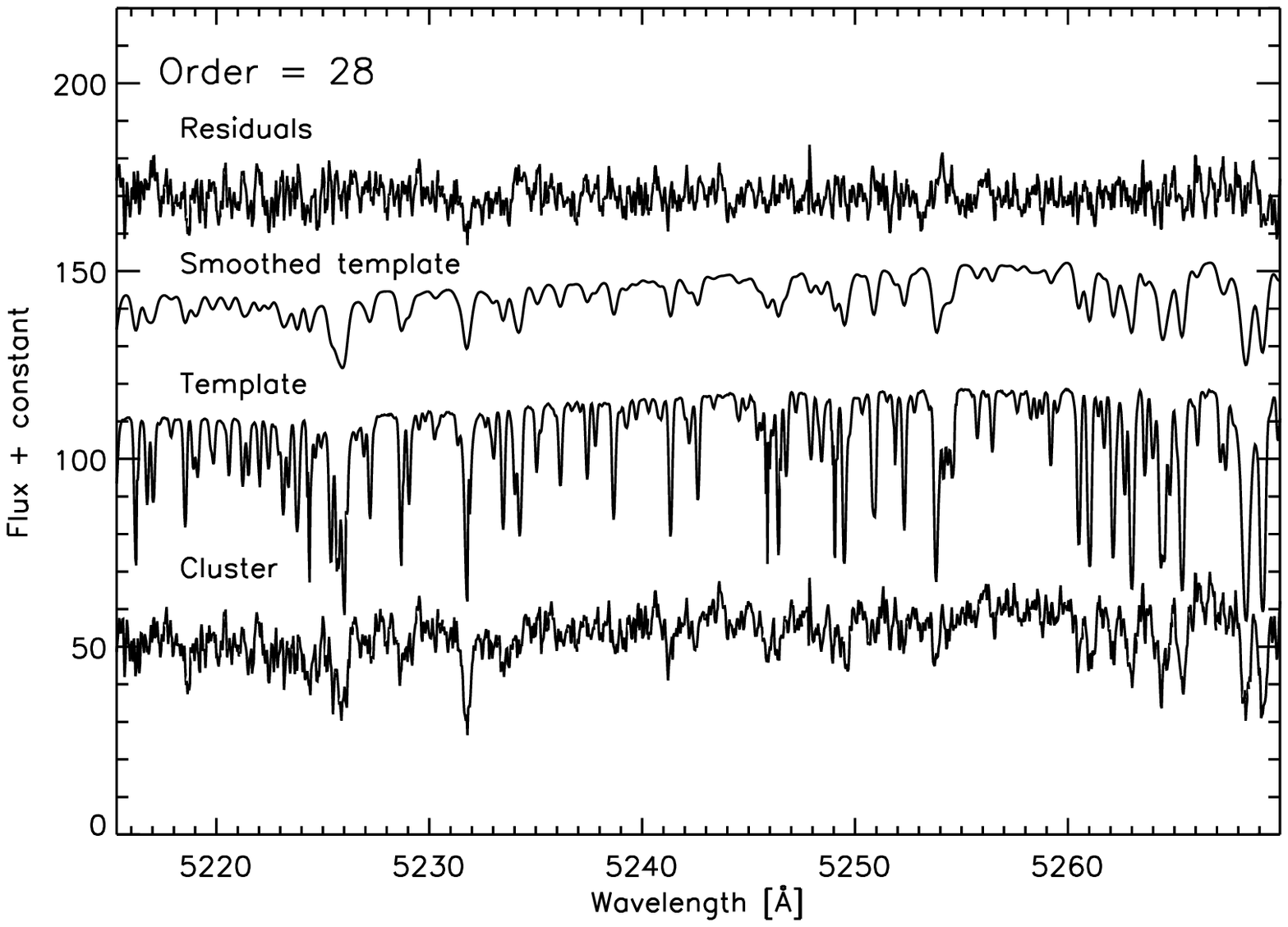}
\end{flushleft}
\figcaption[Larsen.fig1a.ps,Larsen.fig1b.ps]{\label{fig:vd}
  Comparison of cluster spectra and template 
  star spectra for two echelle orders of the cluster M33-H38. All spectra
  are normalized in the same way, but zero-point offsets have been
  applied for clarity. Order 27 includes the region around the Mg$b$ triplet.}
\newpage

\epsfxsize=15cm
\epsfbox{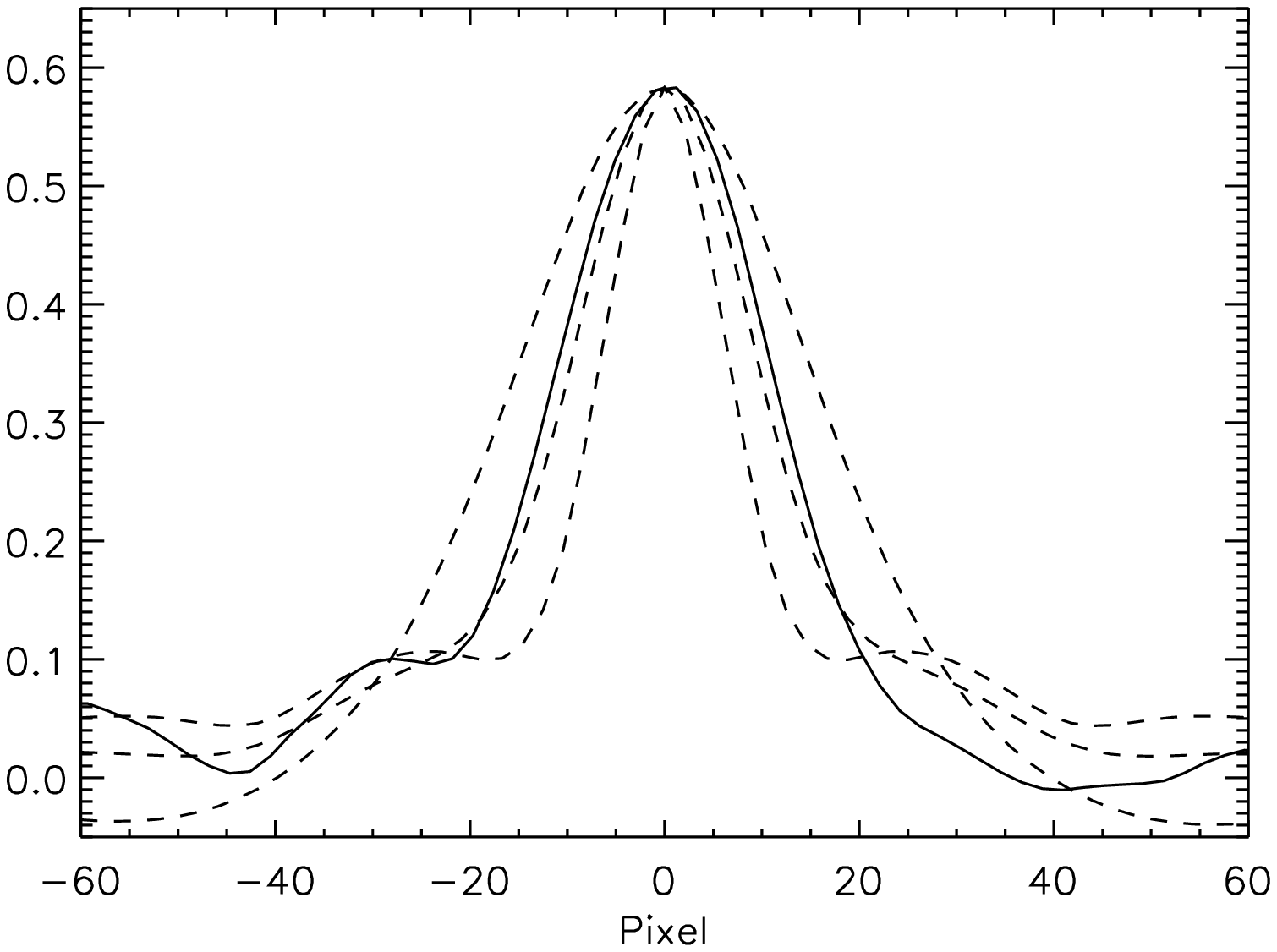}
\figcaption[Larsen.fig2.ps]{\label{fig:ccf}
  Cross-correlation for template star vs.\
  cluster H38 (solid line) and template star broadened with Gaussian
  profiles corresponding to $\sigma_x = 0$, 6 km/s and 12 km/s (dashed
  lines).}
\newpage

\noindent
\begin{minipage}{17cm}
\begin{minipage}{84mm}
\epsfxsize=82mm
\epsfbox{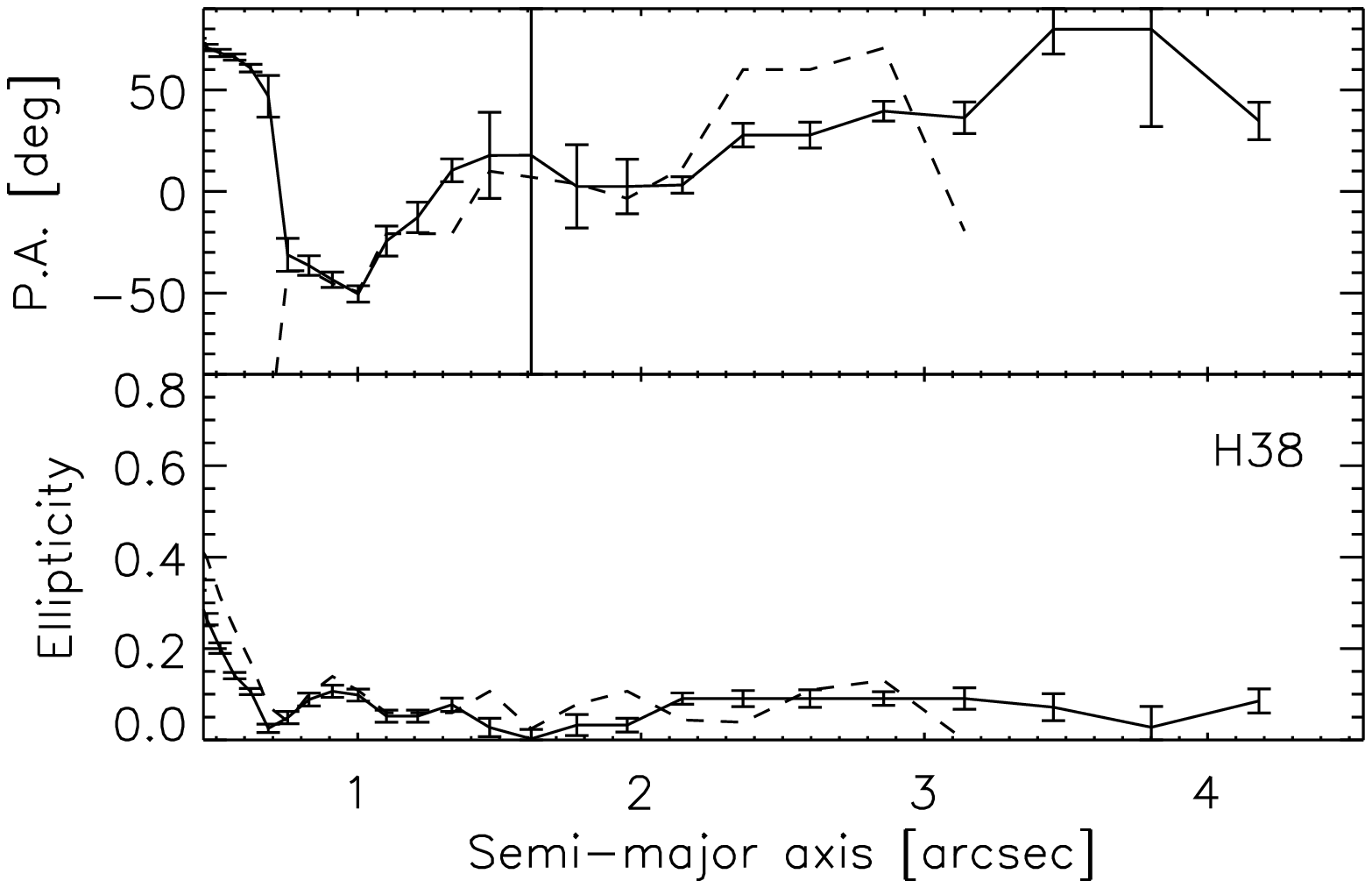}
\end{minipage}
\begin{minipage}{84mm}
\epsfxsize=82mm
\epsfbox{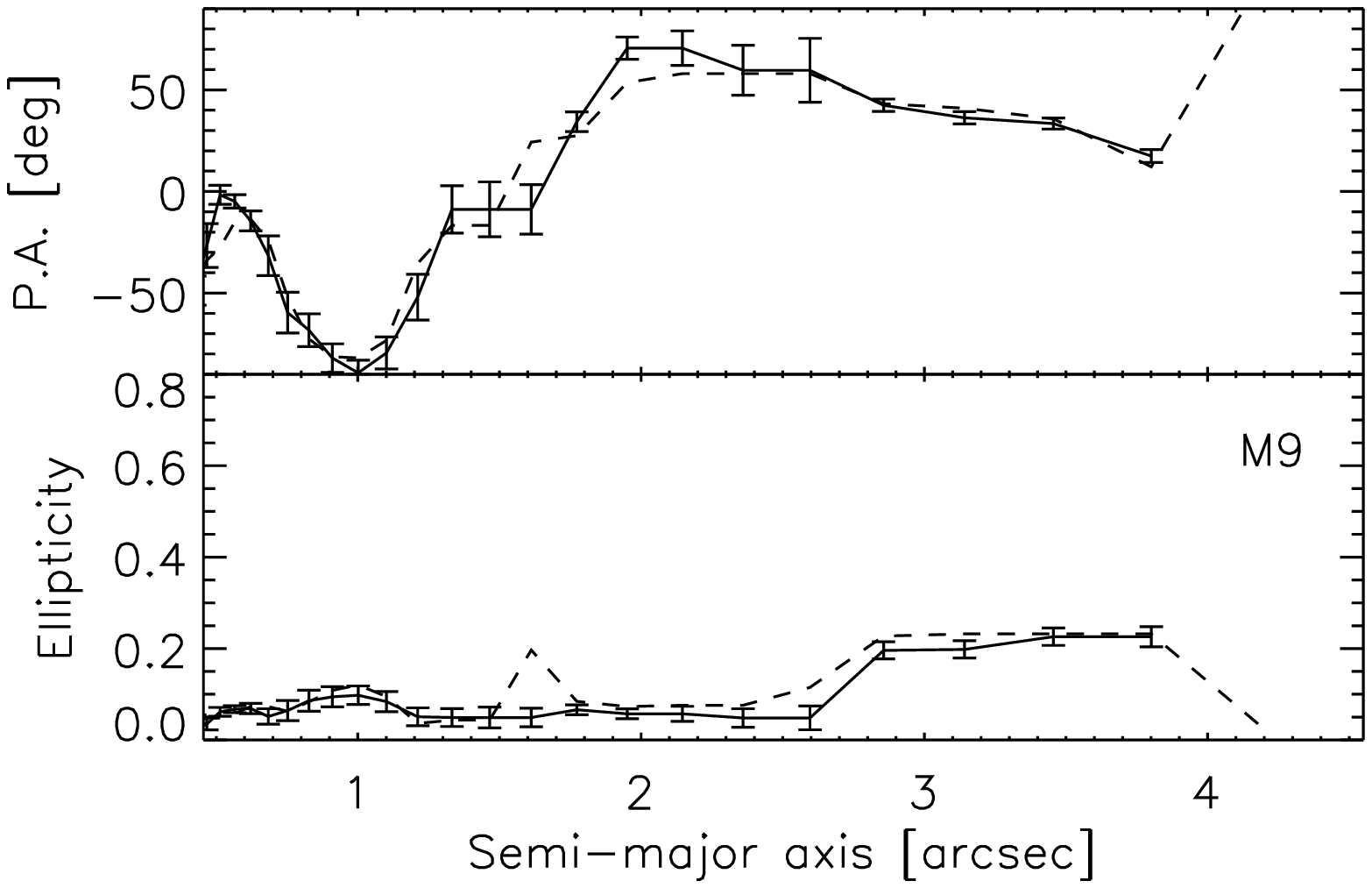}
\end{minipage}
\end{minipage}
\\
\noindent \begin{minipage}{17cm}
\begin{minipage}{84mm}
\epsfxsize=82mm
\epsfbox{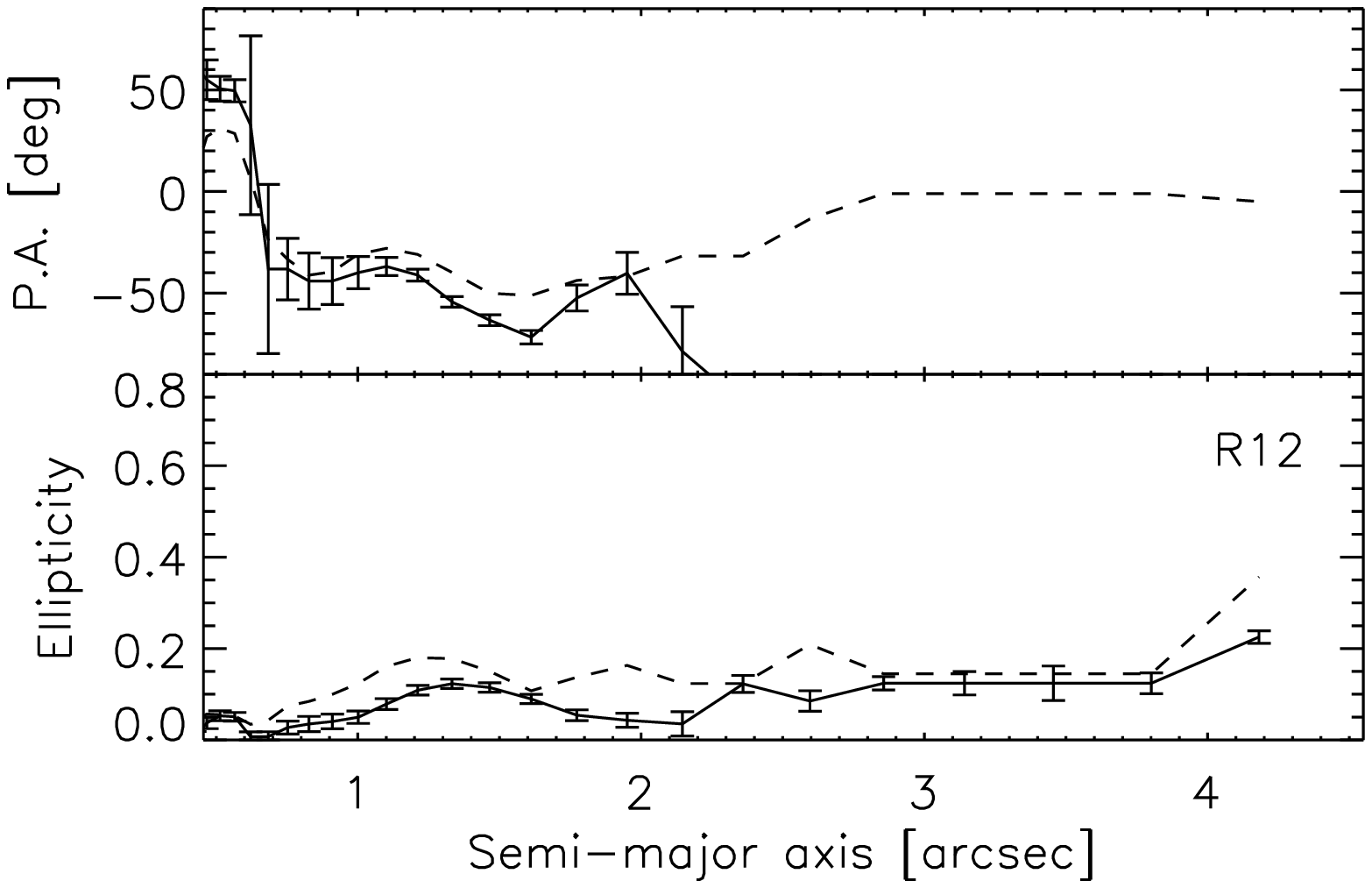}
\end{minipage}
\begin{minipage}{84mm}
\epsfxsize=82mm
\epsfbox{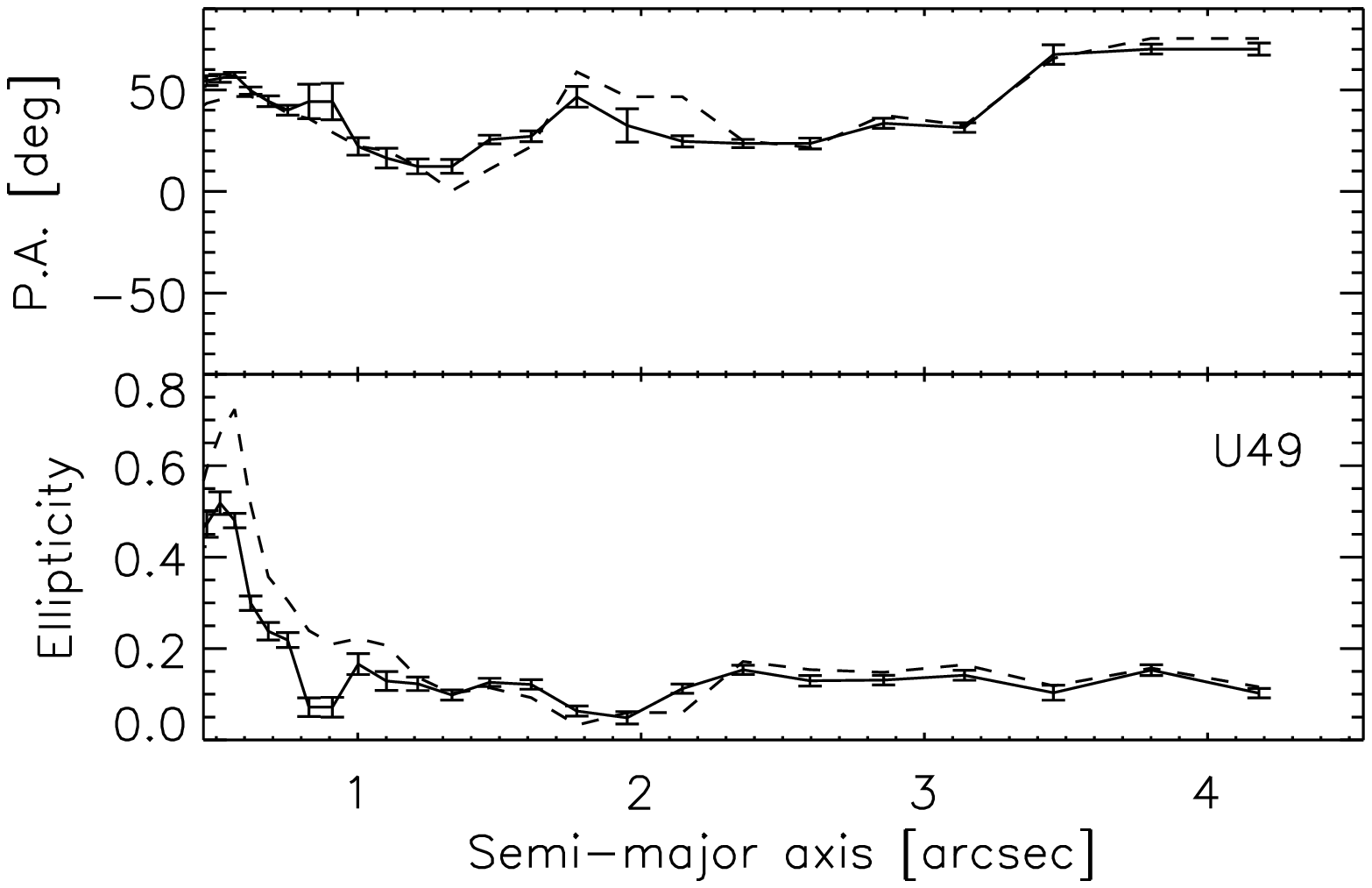}
\end{minipage}
\end{minipage}
\figcaption[Larsen.fig3a.ps,Larsen.fig3b.ps,Larsen.fig3c.ps,Larsen.fig3d.ps]
{\label{fig:efit}Ellipticities and major axis position angles
 for the four globular clusters. Solid and dashed lines indicate fits to
 F555W and F814W images, respectively. Position angles are counted
 N through E.}
\newpage

\epsfxsize=14.5cm
\epsfbox{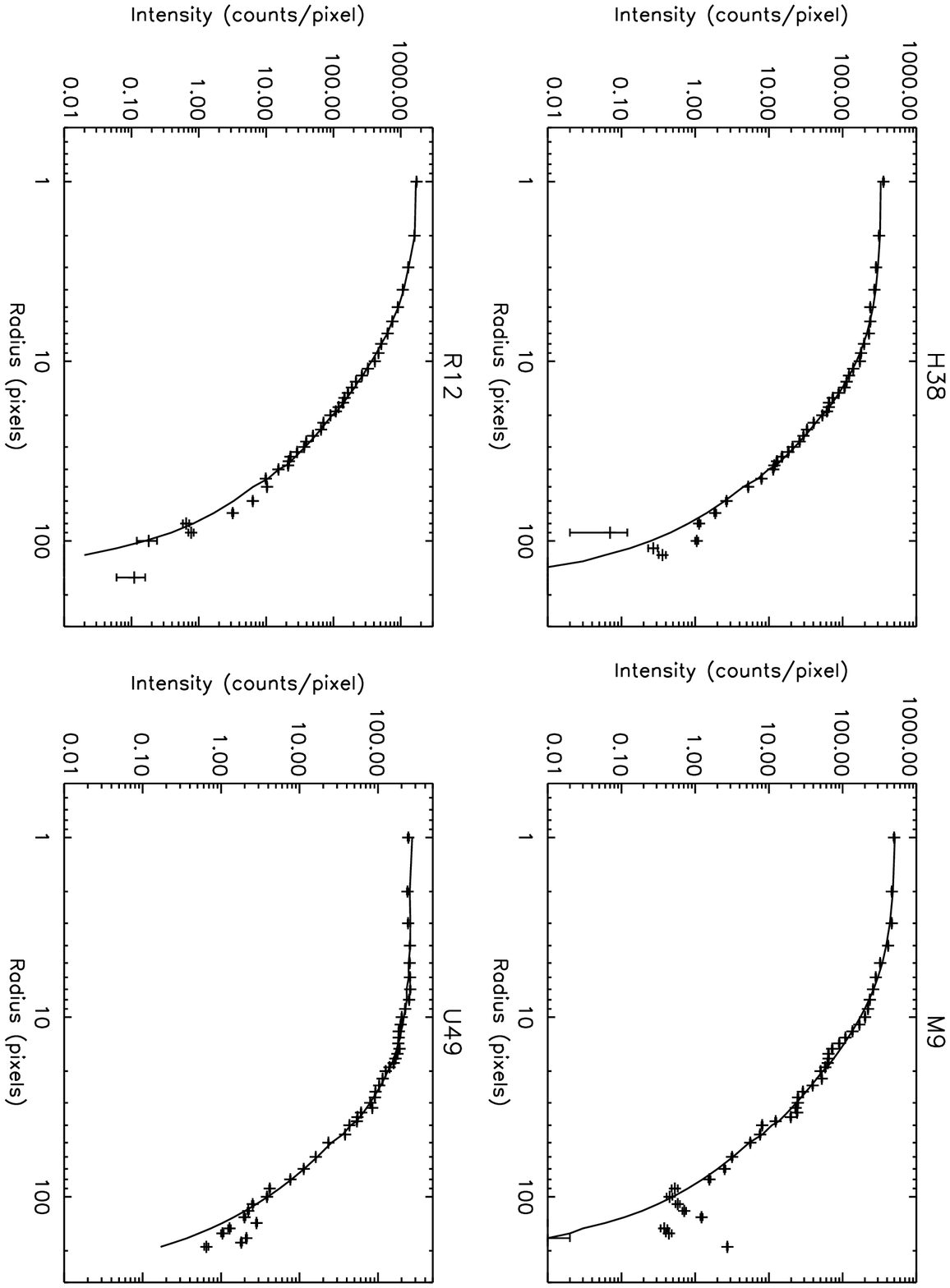}
\figcaption[Larsen.fig4.ps]{\label{fig:kfitv}
  Surface brightness profiles and King profile 
  fits for HST images of the 4 globular clusters in M33.}
\newpage

\epsfxsize=15cm
\epsfbox{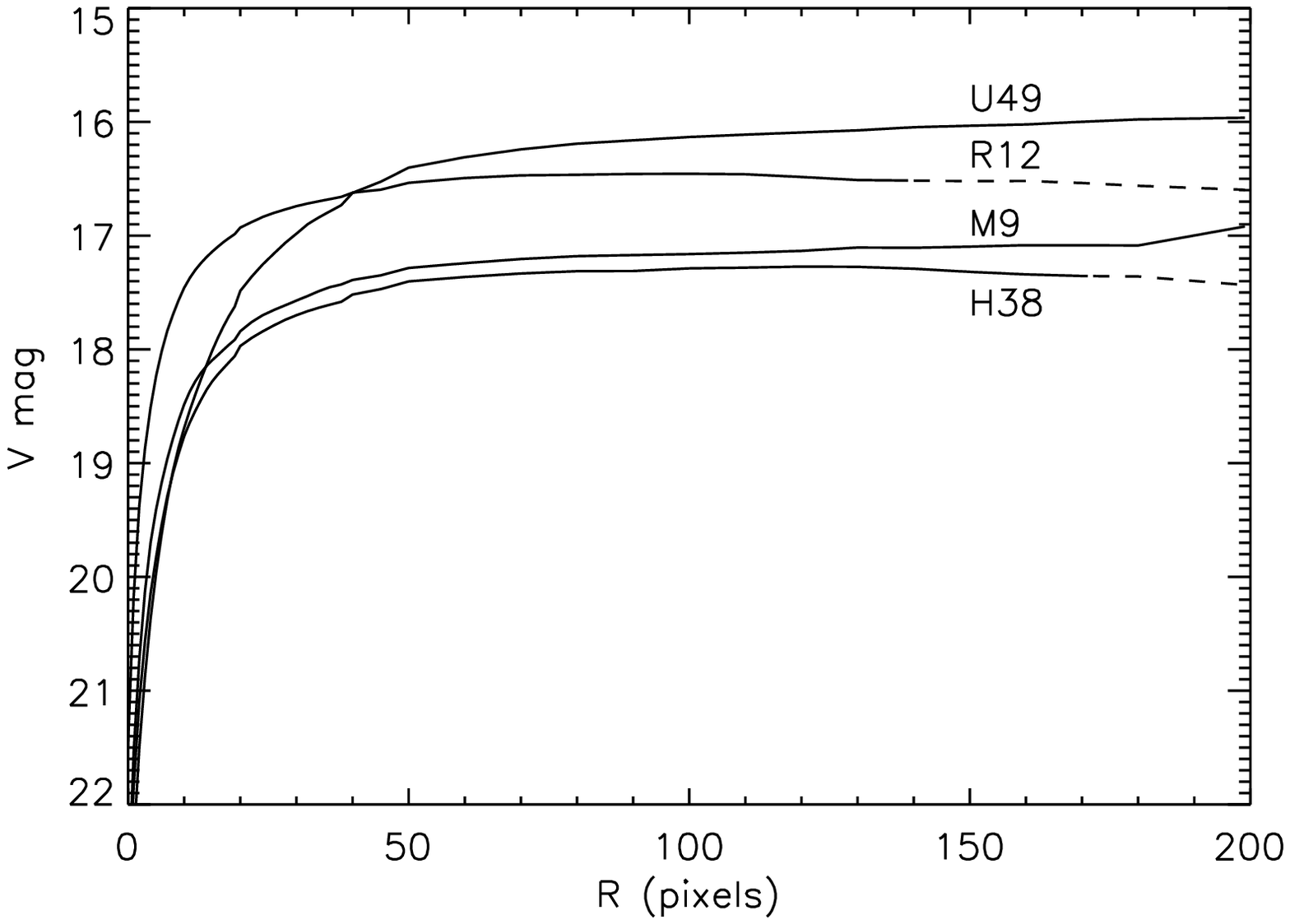}
\figcaption[Larsen.fig5.ps]{\label{fig:cgrow}
  Curves-of-growth for the globular clusters.
  The curves are drawn with dashed lines beyond the tidal radii inferred
  from the King model fits.
}
\newpage

\epsfxsize=15cm
\epsfbox{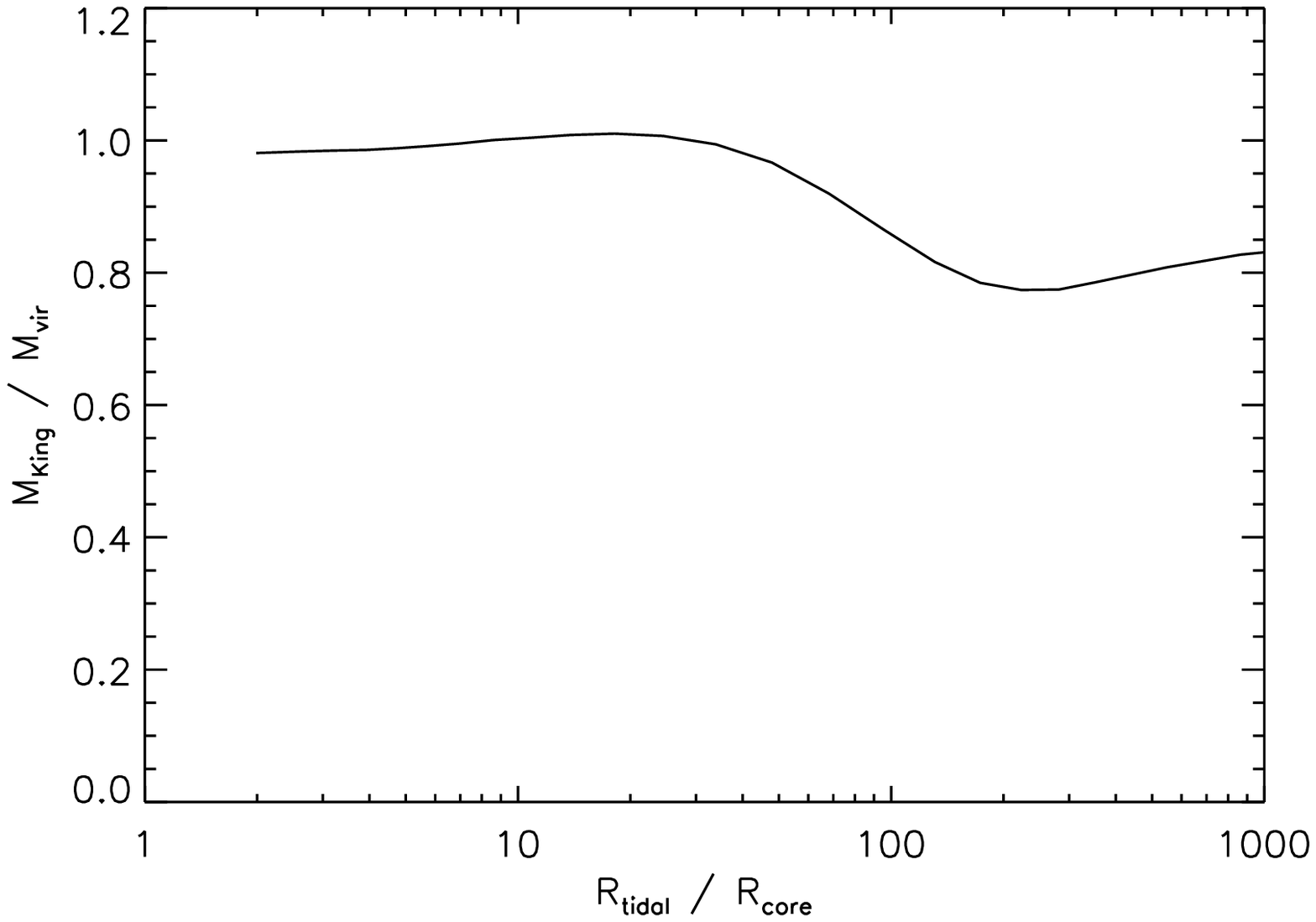}
\figcaption[Larsen.fig6.ps]{\label{fig:mcmp}
  Ratio of King mass to virial mass as a function
  of concentration parameter.}
\newpage

\noindent \begin{minipage}{17cm}
\begin{minipage}{84mm}
\epsfxsize=84mm
\epsfbox[80 370 545 700]{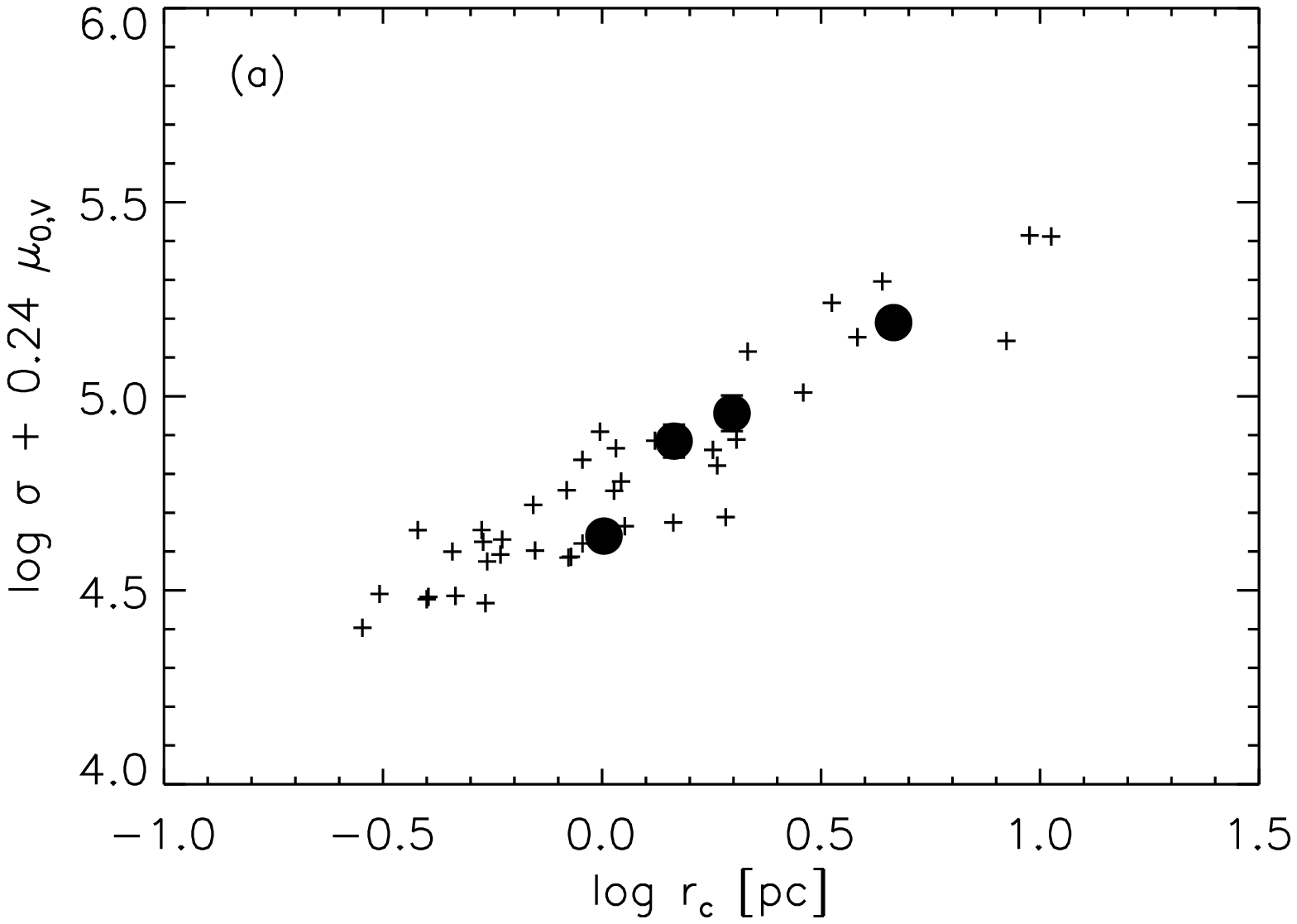}
\end{minipage}
\begin{minipage}{84mm}
\epsfxsize=84mm
\epsfbox[80 370 545 700]{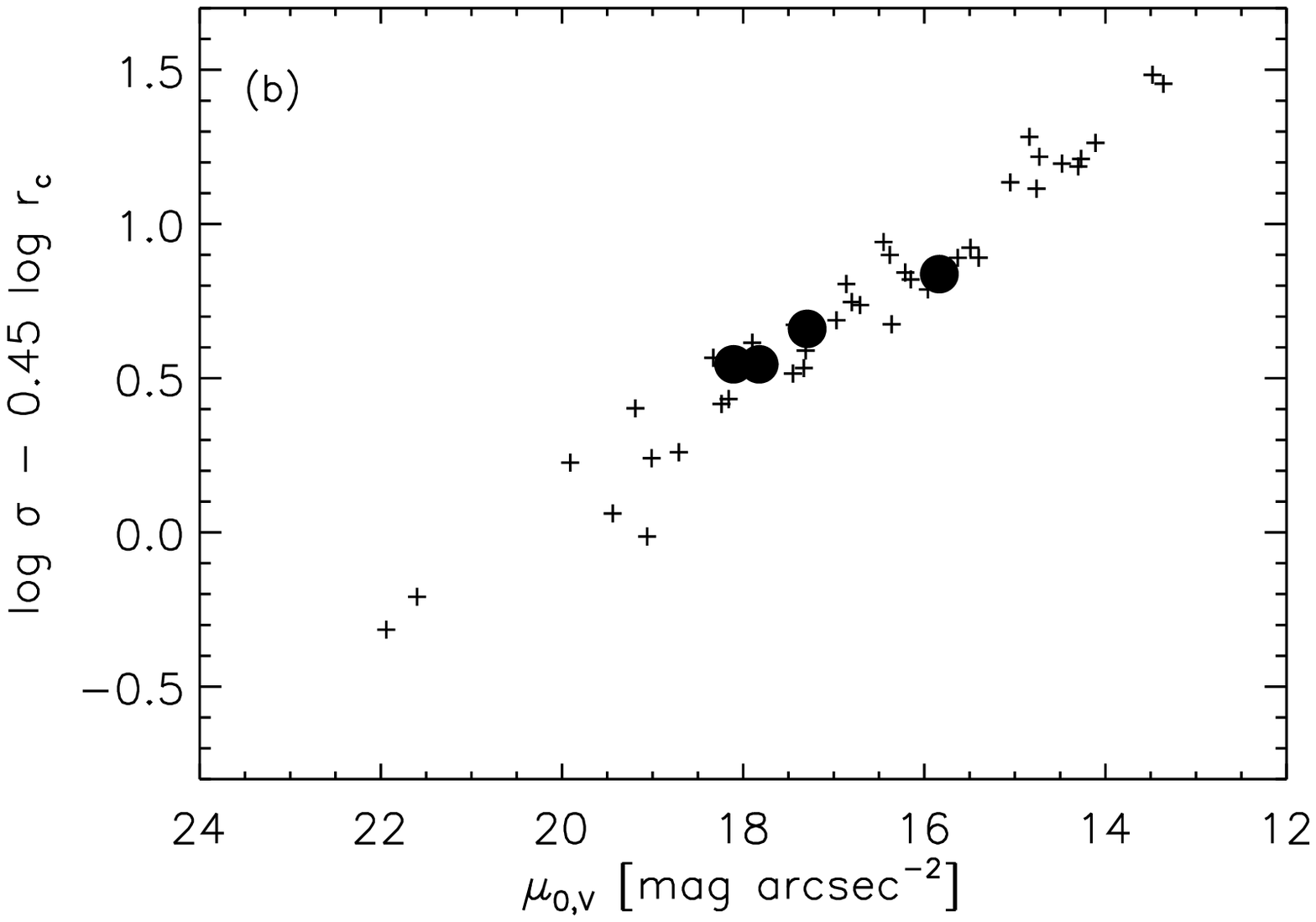}
\end{minipage}
\begin{minipage}{84mm}
\epsfxsize=84mm
\epsfbox[80 370 545 700]{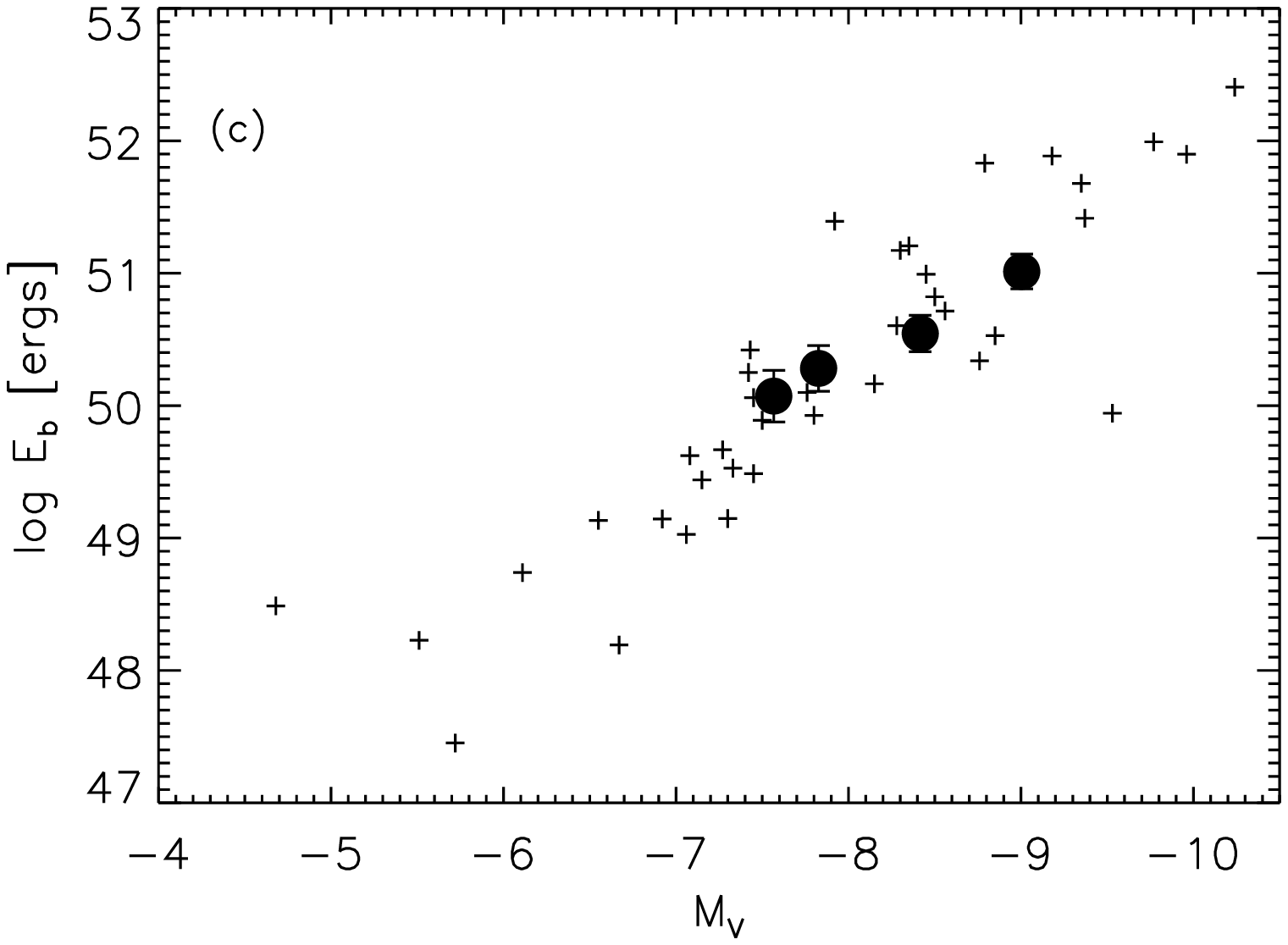}
\end{minipage}
\end{minipage}
\figcaption[Larsen.fig7a.ps,Larsen.fig7b.ps,Larsen.fig7c.ps]{\label{fig:fp}
  Fundamental plane correlations for Milky Way 
 ($+$ symbols) and M33 globular clusters (filled circles).}
\newpage

\begin{deluxetable}{lcccccccccc}
\tablecaption{\label{tab:rv}Radial velocities and measured velocity 
dispersions.}
\tablecomments{$N$ is the number of echelle orders used. The measured
line-of-sight velocity dispersions, $\sigma_x$, are in km/s.  For 
comparison we have included radial velocities from \citet{schom91} and 
\citet{chan02}.}
\tablehead{ 
  ID  &  $N$  &   $v_r$   &  $v_r$ (S91) & $v_r$ (C02) & $\sigma_x$  & rms & err &  $\sigma_x$  & rms & err \\
      &       &   km/s    &  km/s  &  km/s & \multicolumn{3}{c}{Direct fit} &
                             \multicolumn{3}{c}{Cross-correlation}  \\
  (1) & (2)   &   (3)     &  (4)         &   (5)       & (6) & (7) &    (8)       & (9) & (10) & (11)
			     }
\startdata
 H38  & 26 & $-195.4\pm0.8$ & $-100\pm100$ & $-254\pm20$ & 5.33  & 1.29 & 0.25 & 4.42 & 0.84 & 0.16 \\
 M9   & 21 & $-202.4\pm1.1$ & $-300\pm40$  & $-213\pm10$ & 5.57  & 0.98 & 0.21 & 5.01 & 1.02 & 0.22 \\
 R12  & 19 & $-218.4\pm0.7$ & $-190\pm40$  & $-218\pm10$ & 6.69  & 0.37 & 0.08 & 6.31 & 0.52 & 0.12 \\
 U49  & 19 & $-149.8\pm1.5$ & $-180\pm40$  & $-157\pm16$ & 7.04  & 0.90 & 0.21 & 6.52 & 1.18 & 0.27 \\
\enddata
\end{deluxetable}

\begin{deluxetable}{lccccc}
\tablecaption{\label{tab:kfit}King model parameters.}
\tablecomments{Concentration parameter $C$ and effective radius \reff\ 
 are derived from $r_c$ and $W$. Central surface 
 brightnesses have not been corrected for reddening.} 
\tablehead{ 
  ID  &  $r_c$     &      $W$      &   $\mu_0$      &   $C$        & \reff  \\
      &   pc       &               & mag arcsec$^{-2}$  &          &  pc     \\
  (1) &  (2)   & (3)  &  (4)       &     (5)       &   (6)   
      }
\startdata
 F555W &              &               &                &              & \\
 H38  & $1.98\pm0.07$ & $5.97\pm0.22$ & $17.90\pm0.03$  & $17.8\pm2.1$ & $3.94\pm0.16$ \\
  M9  & $1.46\pm0.10$ & $6.74\pm0.38$ & $17.37\pm0.06$ & $28.3\pm6.3$ & $3.88\pm0.38$ \\
 R12  & $1.01\pm0.02$ & $6.76\pm0.12$ & $15.93\pm0.02$ & $28.4\pm2.2$ & $2.67\pm0.10$ \\
 U49  & $4.63\pm0.20$ & $5.40\pm0.28$ & $18.24\pm0.03$ & $13.1\pm1.9$ & $7.81\pm0.33$ \\
 F814W &              &               &                &              & \\
 H38  & $2.09\pm0.13$ & $5.44\pm0.40$ & $16.78\pm0.05$ & $13.6\pm2.4$ & $3.57\pm0.18$ \\
 M9   & $1.54\pm0.07$ & $6.70\pm0.26$ & $16.42\pm0.04$ & $27.7\pm4.8$ & $3.99\pm0.34$ \\
 R12  & $1.11\pm0.03$ & $6.43\pm0.13$ & $14.85\pm0.03$ & $23.1\pm1.8$ & $2.59\pm0.07$ \\
 U49  & $4.48\pm0.18$ & $5.53\pm0.26$ & $17.18\pm0.03$ & $14.0\pm1.8$ & $7.83\pm0.28$ \\
\enddata
\end{deluxetable}

\begin{deluxetable}{lcccccccccc}
\tablecaption{\label{tab:phot}Integrated photometry for the clusters. No
  corrections for reddening have been applied.}
\tablehead{
  ID  & \multicolumn{2}{c}{CS88} & \multicolumn{2}{c}{CBF01} & 
       \multicolumn{6}{c}{This work} \\ 
  \cline{6-11}
      &      &      &      &       & \multicolumn{2}{c}{$r$ as in CBF01} &
                                     \multicolumn{2}{c}{$r$ as in CS88} &
				     \multicolumn{2}{c}{$r$ = tidal} \\
      & $V$  & \vi  & $V$  &  \vi  &  $V$  & \vi & $V$  & \vi &  $V$  & \vi \\
  (1) & (2)  & (3)  & (4)  &  (5)  &  (6)  & (7) & (8)  & (9) & (10)  & (11)
}
\startdata
H38 & 17.26 & ...  & 17.25 & 1.07  & 17.43 & 1.11 & 17.31 & 1.07 & 17.35 & 1.09\\
M9  & 17.17 & 1.06 & 17.16 & 1.02  & 17.33 & 1.05 & 17.18 & 1.06 & 17.09 & 1.06\\
R12 & 16.38 & 1.19 & 16.37 & 1.15  & 16.55 & 1.17 & 16.46 & 1.17 & 16.52 & 1.19\\
U49 & 16.24 & 1.17 & 16.23 & 1.03  & 16.43 & 1.03 & 16.19 & 1.03 & 15.97 & 1.00\\
\enddata
\end{deluxetable}

\begin{deluxetable}{lccccccc}
\tablecaption{\label{tab:mass}Mass estimates, central densities and
  mass-to-light ratios.}
\tablecomments{$\sigma_0$ and $\sigma_{\infty}$ are the central and global
1-d velocity dispersions, derived from the observed velocity dispersions
$\sigma_x$ in Table~\ref{tab:rv} based on the cross-correlation technique.}
\tablehead{
  ID & $\sigma_0$ & $\sigma_{\infty}$ & $M_{\rm king}$ & $M_{\rm vir}$ 
    & $\rho_0$ & (M/L)$_V$ & $E_b$ \\
     &  km/s      &     km/s        & \multicolumn{2}{c}{$\times10^5\,\msun$}
              &  $\msun \, {\rm pc}^{-3}$  &  & $\times10^{50}$ ergs \\
  (1) & (2)       & (3)               &  (4)           &  (5) & (6) & (7) & (8)
    }
\startdata
 H38 & 4.77       &    3.99         &  $1.46\pm0.33$ & $1.44\pm0.33$ & $1200\pm300$  & $1.65\pm0.37$  & $1.18\pm0.53$ \\
 M9  & 5.42       &    4.46         &  $1.77\pm0.35$ & $1.77\pm0.35$ & $2500\pm700$  & $1.57\pm0.31$  & $1.91\pm0.76$ \\
 R12 & 6.91       &    5.69         &  $2.00\pm0.32$ & $2.00\pm0.32$ & $8600\pm1500$ & $1.03\pm0.16$  & $3.50\pm1.11$ \\
 U49 & 6.99       &    5.88         &  $6.23\pm0.96$ & $6.20\pm0.95$ & $ 460\pm90$   & $1.87\pm0.29$  & $10.3\pm3.1$ \\
\enddata
\end{deluxetable}

\end{document}